\documentclass[]{article}

\usepackage{graphicx}
\usepackage{subfig}
\usepackage{epstopdf}
\usepackage{epsfig}
\usepackage{authblk}

\usepackage[T1]{fontenc}
\usepackage[utf8]{inputenc}

\title{The Be Phenomenon in the Mono-Periodic STEREO Star; 13\,Tau}
\author[1,2]{D.Ozuyar\thanks{dozuyar@ankara.edu.tr}}
\author[2]{I. R. Stevens\thanks{irs@star.sr.bham.ac.uk}}

\affil[1]{Ankara University, Astronomy and Space Sci. Dept., Tandogan, Ankara, 06100, Turkey}
\affil[2]{The University of Birmingham, School of Physics and Astronomy, Birmingham, B15 2TT, UK}

\begin{document}

\maketitle

\begin{abstract}
13\,Tau is a rarely studied bright B9V type star ({\sl V} = 5.68 mag), which shows a weak and double-peaked H{$\alpha$} emission profile in its spectra. In this study, we presented high-precision photometric data of 13\,Tau taken by the {\sl STEREO} satellite between 2007 and 2011, and compared the results to the spectroscopic findings to shed light on the Be phenomenon in the star. From the frequency analysis of the five-year data, we detected that 13\,Tau has exhibited a mono-periodic light variation ($f$ = 1.80487(1) cd$^{-1}$; A $\sim$ 2.76(8) mmag). The analysis revealed that frequency and amplitude values of the seasonal light curves varied from one year to another. From the spectroscopic data, we figured out that the equivalent widths of the H{$\alpha$} lines also showed variability, which seemed connected to the changes seen in both frequency and amplitude. 
\end{abstract}

\section{Introduction}

Classical Be stars are single, rapidly rotating (250-500 kms$^{-1}$ \cite{Struve1931}) and non-radially pulsating B stars \cite{rivinius2013}. They are surrounded by an outwardly diffusing Keplerian disk, whose evolution is governed by viscosity \cite{rivinius2016}. This equatorial disk is the source of the emission lines observed in the spectrum. Variability in the line profile is a common situation and can continue from weeks to decades. While long-term profile variations are attributed to a stable one-armed density wave pattern in the disk, variabilities on the time-scales of a few days are due to the close circumstellar environment or due to the stellar surface \cite{rivinius2013}. The stellar rotation, pulsation periods, Keplerian orbital period and viscous transport times through the inner disk also fall in this range \cite{rivinius2013}. Additionally, these stars exhibit modulations in their light curves (LCs). It is known that non-radial pulsations (NRPs) and rotation of the central star are responsible for these periodicities. These mechanisms are also believed to be related to the ejected material from the stellar surface to form a Keplerian disk \cite{rivinius2016}.

In this context, the photometric and spectroscopic data of 13\,Tau are analyzed in terms of the Be phenomenon in this study. 13\,Tau is a rarely studied bright B9V type classical Be star, which shows a weak and double-peaked H{$\alpha$} emission profile in its spectrum \cite{Slettebak1982}. Based on the analysis of the {\sl Kepler~K2} data, Balona \cite{Balona2016} gives its rotational period to be 0.555(1) days and also reports a visible harmonic in the data.

\section{Data Analysis}

The high-precision photometric data of 13 Tau are obtained from the HI-1A camera of the {\sl STEREO} satellite between 2007 and 2011. Seasonal data comprised an observation interval of ~20 days. The cadence of each data chunk is 40 minutes and the Nyquist frequency is $\sim$18 cd$^{-1}$. A more detailed description of the data preparation and background information can be found in Sangaralingam \& Stevens \cite{ss2011} and Whittaker et al. \cite{Whittaker2013}. Also, the details of the HI instruments can be accessed from the paper of Eyles \cite{Eyles2009}. Five years of annual and combined time series were analyzed with the Lomb-Scargle (LS) algorithm. Frequencies whose amplitudes were greater than the significance level having 99\% probability were determined. The detection precision in five-year combined data was around 10$^{-5}$ cd$^{-1}$ in frequency and 10$^{-4}$ mag in amplitude. The photometric data were supported with 13 high-resolution H{$\alpha$} observations taken from the Be Star Spectra (BESS) Database\footnote{http://basebe.obspm.fr/basebe/}. These spectra were derived from 10 different observation sites by using 11 different equipments in France and Spain between 2006 and 2012. During the analysis, heliocentric velocity corrections were applied, and telluric lines were removed by using a reference spectrum. For the continuum normalization, the margin of error was considered as 3\%\ \cite{Jones2011}. For the calculation of equivalent widths (EWs), SPLOT package of the IRAF was used. 
  
 \begin{table}[!t]
\footnotesize
\begin{center}
\caption[Frequency analysis results of 13\,Tau]{Frequency analysis results of the seasonal and combined data sets. }
\smallskip
\begin{tabular}{l r l r r r}
\hline
\noalign{\smallskip}
\textbf{Year}&\textbf{Mid-Obs.Time}&\textbf{No}&\textbf{Frequency}&\textbf{Amplitude}&\textbf{SNR}\\
&\textbf{(HJD)}&\textbf{\#}&\textbf{(cd$^{-1}$)}&\textbf{(mmag)}&\\
\noalign{\smallskip}
\hline
\noalign{\smallskip}
& Combined	&	$f_1$	&	1.80487(1)	&	2.76(08)	&	13.40		\\
\hline
\noalign{\smallskip}
2007 & 2454221	&	$f_1$	&	1.794(2)	&	2.24(17)	&	5.91		\\
\hline
\noalign{\smallskip}
2008 & 2454566	&	$f_1$	&	1.803(2)	&	2.67(15)	&	6.50		\\
\hline
\noalign{\smallskip}
2009 & 2454910	&	$f_1$	&	1.809(2)	&	2.72(16)	&	6.77		\\
\hline
\noalign{\smallskip}
2010 & 2455255	&	$f_1$	&	1.808(2)	&	2.81(18)	&	6.88		\\
\hline
\noalign{\smallskip}
2011 & 2455599	&	$f_1$	&	1.806(2)	&	3.55(25)	&	6.48		\\
\hline
\end{tabular}
\label{tab:Table1}
\end{center}
\end{table}

\section{Results and Discussion}

\begin{figure}[!t]
\begin{center}
\includegraphics[width=1.\textwidth]{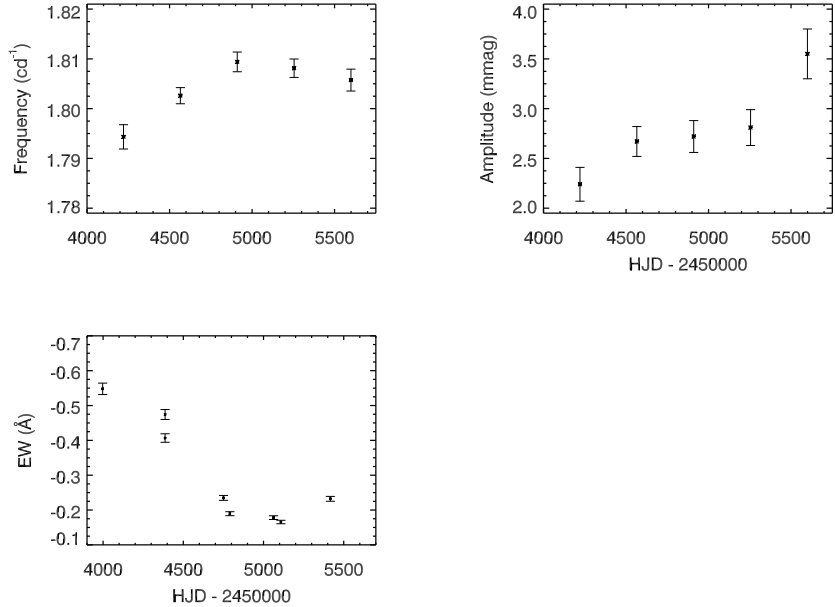}
\caption{Seasonal frequency, amplitude and equivalent width variations of 13\,Tau.}
\label{fig:figure1}
\end{center}
\end{figure}

From the LS analysis of the five-year combined data, we detected that the star has exhibited mono-periodic variations ($f$ = 1.80487(1) cd$^{-1}$ and A = 2.76(8) mmag). The seasonal LCs indicated that the main frequency and its amplitude varied over five years as seen in Table~\ref{tab:Table1}. From Fig.~\ref{fig:figure1}, the variation had a decreasing trend, where the frequency rapidly shifted from 1.794 to 1.809 cd$^{-1}$ between HJD2454221 and HJD2454910, and then moderately decreased to 1.806 cd$^{-1}$ on HJD2455599. In addition, although the amplitude appeared to be constant (A $\sim$ 2.73 mmag) between HJD2454566 and HJD2455255, five years of data showed an increase in the amplitude intensity.

The individual spectra of 13\,Tau exhibited a deeper central absorption line surrounded by a double-peaked emission profile as well as a shallower broad absorption line. After the removal of the broad absorptions, we calculated the EW values and observed that the data exhibited a decreasing profile of about 0.6 \AA\ (Figure~\ref{fig:figure1}); the EWs increased (in other words the emission strength decreased) between HJD2454000 and HJD2454700 whereas they remained constant from HJD2454700 to HJD2455700. This overall EW variation seemed to be inversely proportional to those observed in both frequency and amplitude. From the decreasing emission strength and increasing frequency amplitude, it could be speculated that the star has been prepared to go into anew outburst phase based on the statement that the pulsation amplitude increases during the outbursts of pulsational Be stars \cite{Kurtz2015}.

Compared to other Be star samples, the amplitude spectrum of 13\,Tau shows no particular frequency group. Considering the mono-periodic distribution, the star seems to be a member of $\lambda$ Eri variables whose photometric variations are dominated by short term periodic processes caused due to circumstellar environment \cite{rivinius2013}. Also, while NRPs are known to be effective in the disk formation of many Be-stars \cite{Rivinius1998,rivinius2013}, there is no interference of such modes in 13\,Tau (at least in {\sl STEREO} photometric observations). The absence of NRPs may be a result of the inclination angle ($i$) of the star. Based on the definitions about the line profile shapes of Be stars given by Hanuschik \cite{hanuschik1996}, the H{$\alpha$} emission line of the star resembles a wine bottle profile, which occurs close to $i = 0^{\circ}$. Therefore, the expected NRPs may not be observed due to this restricted line of sight. Additionally, the detection of NRPs may be limited by the threshold used for the analysis. On the other hand, according to Rivinius \cite{rivinius2013}, at a high $W$, ($W=v_{rot}/v_{orb}$, where $v_{rot}$ and $v_{orb}$ are the rotational and Keplerian circular orbital velocities at the equator, respectively) a weak mechanism such as single pulsation may be effective in the disk formation. Otherwise, it is traditionally believed that a single pulsation does not provide sufficient kinetic energy and angular momentum to form a rotationally-supported circumstellar disk \cite{Owocki2004}. This means that there may be some other mechanisms that trigger the mass loss and interact with the disk structure.

The {\sl STEREO} gives some hints about the variability seen in 13\,Tau by providing uninterrupted observations and high-precision measurements. However, the current data are not sufficient to study on the disk formation and the origin of the photometric variabilities. Also, there is no archival data related to the star. Therefore, long-term photometric and spectroscopic follow-ups are needed in order to unravel the nature and evolution status of 13\,Tau.

\section*{Acknowledgements}

We acknowledge assistance from Sangaralingam, V. and Whittaker, G. in the production of the data used in this study.

This work has made use of the BeSS database, operated at LESIA,
Observatoire de Meudon, France: http://basebe.obspm.fr

The STEREO Heliospheric imager was developed by a collaboration that
included the Rutherford Appleton Laboratory and the University of
Birmingham, both in the United Kingdom, and the Centre Spatial
de Lige (CSL), Belgium, and the US Naval Research Laboratory
(NRL),Washington DC, USA. The STEREO/SECCHI project is an
international consortium of the Naval Research Laboratory (USA),
Lockheed Martin Solar and Astrophysics Lab (USA), NASA Goddard 
Space Flight Center (USA), Rutherford Appleton Laboratory (UK), University of Birmingham (UK), Max-Planck-Institut fr
Sonnen-systemforschung (Germany), Centre Spatial de Lige (Belgium), Institut dOptique Thorique et Applique (France) and Institut dAstrophysique Spatiale (France). This research has made use of
the SIMBAD data base, opened at CDS, Strasbourg, France. 

This research has also made use of NASA's Astrophysics Data System.

\end{document}